\documentclass{elsart}

\usepackage{epsfig}
\usepackage{longtable}

\usepackage{amsmath}
\usepackage{amssymb}

\begin{document}

\begin{frontmatter}



\title{Rotational spectroscopy of isotopic vinyl cyanide, H$_2$C=CH$-$C$\equiv$N, 
in the laboratory and in space\thanksref{Widmung}}
\thanks[Widmung]{This work is dedicated to Edward A. Cohen and Herbert M. Pickett 
in recognition of their many contributions to the theory and application of 
molecular spectroscopy.}


\author[Koeln,Bonn]{Holger~S.P. M\"uller\corauthref{cor}},
\ead{hspm@ph1.uni-koeln.de}
\corauth[cor]{Corresponding author.}
\author[Bonn]{Arnaud Belloche},
\ead{belloche@mpifr-bonn.mpg.de}
\author[Bonn]{Karl M. Menten},
\ead{menten@mpifr-bonn.mpg.de}
\author[Bonn]{Claudia Comito},
\ead{ccomito@mpifr-bonn.mpg.de}
\author[Bonn]{Peter Schilke},
\ead{schilke@mpifr-bonn.mpg.de}

\address[Koeln]{I.~Physikalisches Institut, Universit\"at zu K\"oln,
  Z\"ulpicher Str. 77, 50937 K\"oln, Germany}
\address[Bonn]{Max-Planck Institut f\"ur Radioastronomie, Auf dem H\"ugel 69, 
53121 Bonn, Germany}

\begin{abstract}
The rotational spectra of singly substituted $^{13}$C and $^{15}$N isotopic species 
of vinyl cyanide have been studied in natural abundances between 64 and 351~GHz. 
In combination with previous results, greatly improved spectroscopic parameters 
have been obtained which in turn helped to identify transitions of the $^{13}$C 
species for the first time in space through a molecular line survey of 
the extremely line-rich interstellar source Sagittarius B2(N) 
in the 3~mm region with some additional observations at 2~mm.
The $^{13}$C species are detected in two compact ($\sim 2.3''$), hot (170~K) 
cores with a column density of $\sim 3.8 \times 10^{16}$ and 
$1.1 \times 10^{16}$ cm$^{-2}$, respectively.  
In the main source, the so-called ``Large Molecule Heimat'', 
we derive an abundance of $2.9 \times 10^{-9}$ for each $^{13}$C species 
relative to H$_2$. An isotopic ratio $^{12}$C/$^{13}$C of 21 has been measured.
Based on a comparison to the column 
densities measured for the $^{13}$C species of ethyl cyanide also detected in 
this survey, it is suggested that the two hot cores of Sgr~B2(N) are in 
different evolutionary stages. 
Supplementary laboratory data for the main isotopic species recorded between 
92 and 342~GHz permitted an improvement of its spectroscopic parameters as 
well.
\end{abstract}

\begin{keyword}
vinyl cyanide \sep acrylonitrile \sep propenenitrile \sep
rotational spectroscopy \sep submillimeter wave spectroscopy \sep 
interstellar molecule \sep star-forming region \sep Sagittarius B2
\end{keyword}
\end{frontmatter}

\section{Introduction}
\label{intro}

Vinyl cyanide, (VyCN for short), also known as acrylonitrile or propenenitrile, 
H$_2$C=CH$-$C$\equiv$N or shorter C$_2$H$_3$CN, is an important molecule in space. 
It was detected in the massive star-forming region Sagittarius B2 (Sgr B2 for short) 
by Gardner and Winnewisser as early as 1975 \cite{VyCN_det_space}. 
Massive star-forming regions are comparatively dense and hot, thus, transitions 
of VyCN in its $v_{11} = 1$ and  $v_{15} = 1$ excited vibrational states have been 
detected in Orion KL \cite{Orion-KL_325-360} and in Sgr B2(N) \cite{VyCN-vib}.
Matthews and Sears reported the first detection of this molecule in a dark cloud 
(TMC-1) \cite{VyCN_TMC-1}; and very recently, Ag\'undez et al. reported its detection 
in the circumstellar envelope of the late-type star IRC~+10216 (CW~Leo) \cite{VyCN_IRC+10216}.

VyCN had also been among the molecules whose rotational spectra were studied 
rather early by Wilcox et al. in 1954 \cite{VyCN_Wilcox_1954} and by 
Costain and Stoicheff in 1959 \cite{VyCN_Costain_1959}. 
The latter also investigated the singly-substituted $^{13}$C species as well 
as C$_2$H$_3$C$^{15}$N and H$_2$CCDCN. 
More recently, additional reports have been published 
\cite{VyCN_Gerry_1973,VyCN_Demaison_1973,VyCN_Sutter_1985,VyCN_Cazzoli_1988,VyCN_Demaison_1994,VyCN_Baskakov_1996,VyCN_Colmont_1997}.
Its dipole moment components \cite{VyCN_Wilcox_1954,VyCN_Sutter_1985} 
as well as increasingly accurate $^{14}$N hyperfine structure parameters 
\cite{VyCN_Costain_1959,VyCN_Sutter_1985,VyCN_Baskakov_1996,VyCN_Colmont_1997} 
have been obtained. A total of 14 isotopic species have been investigated in 
Ref~\cite{VyCN_Colmont_1997}, including all stable singly substituted ones, 
all multiply deuterated ones except D$_2$CCHCN, and the singly substituted 
$^{13}$C species of H$_2$CCDCN. The rotational constants were used to derive 
structural parameters according to several structural models.
Finally, a combined fit of these previous data was reported by Thorwirth et al. 
in their work on the related vinyl acetylene molecule \cite{VA_2004}.

We have carried out millimeter line observations of Sgr B2(N)
with the IRAM 30\,m telescope. These observations yielded a variety of 
interesting results, e.g. the first detection of aminoacetonitrile, 
a possible precursor of the simplest aminoacide glycine, which was secured 
by accompanying interferometric observations \cite{AAN-det}.
In the course of the ongoing analyses of the observational data supplementary 
laboratory spectroscopic investigations have been carried out to facilitate 
these and future analyses.

In the present contribution we report on laboratory rotational spectroscopic 
investigations of the singly substituted isotopic species of VyCN containing 
$^{13}$C or $^{15}$N as well as supplementary data on the main isotopic species. 
In addition, we present the first detection of $^{13}$C containing isotopologues 
of VyCN in space.

\section{Experimental details}
\label{exp}

Conventional absorption spectroscopy was employed in Cologne for the 
laboratory investigations. 
The measurements were carried out in a 3~m long glass cell at
room temperature in a static mode. A commercial C$_2$H$_3$CN sample was 
used at pressures of less than 1~Pa for stronger lines and around 2~Pa for 
the generally weak lines. 
Two commercial millimeter-wave synthesizers AM-MSP~1 and 2 (Analytik \& 
Me{\ss}technik GmbH, Chemnitz, Germany) covered the 4 and 3~mm frequency ranges 
(52$-$119\,GHz); because of the detector cut-off, the lowest measured line was 
near 64~GHz.
Additional measurements have been performed in the frequency range 239$-$351~GHz 
with the Cologne Terahertz Spectrometer \cite{CTS1994} using a phase-locked 
backward-wave oscillator as source.
A liquid He-cooled InSb hot-electron bolometer was used as detector throughout 
the investigations.

The uncertainties of the lines were estimated by judging the
line shape and the signal-to-noise ratios.
Investigations, e.\,g., into rotational spectra of SO$_2$~\cite{SO2_new}
and SiS~\cite{SiS_new} demonstrated that relative uncertainties 
$\Delta\nu/\nu$ of 10$^{-8}$ and better can be achieved up to and even 
beyond 1~THz in our laboratory not only for strong, but also for
fairly weak, isolated lines.

\section{Observed laboratory spectra and analysis}
\label{obs}

\begin{figure}[ht]
   \begin{center}
   \includegraphics[width=12cm]{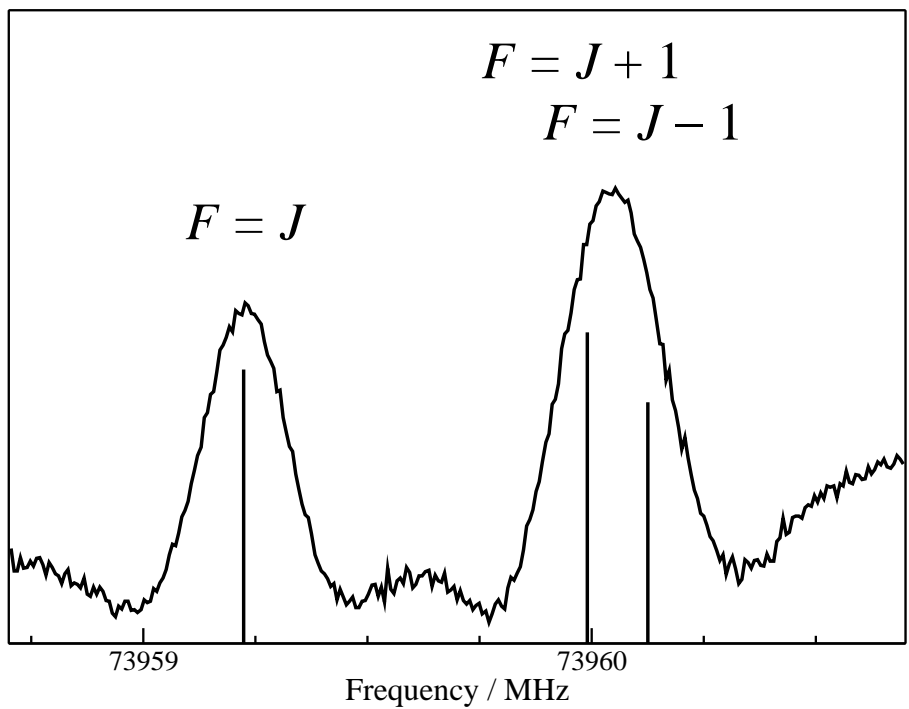}
   \caption{Portion of the rotational spectrum of vinyl cyanide in the region 
            of the prolate paired $J_{K_a,K_c} = 8_{7,1} - 7_{7,0} \ \& \ 8_{7,2} - 7_{7,1}$ 
            transition of H$_2^{13}$C=CH$-$C$\equiv$N demonstrating the partially resolved 
            $^{14}$N hyperfine splitting. The calculated positions and relative intensities 
            of the individual hyperfine components are indicated by vertical lines.
            \label{VyCN_lab-fig}}
   \end{center}
\end{figure}

Vinyl cyanide is a planar asymmetric top that is, with $\kappa = -0.9798$ 
for the main isotopic species in its ground vibrational state, quite close 
to the prolate symmetric limit of $-1$. It has a large $a$-dipole moment 
component of 3.815~(12)~D and a considerably smaller, but still sizeable 
$b$-dipole moment component of 0.894~(68)~D \cite{VyCN_Sutter_1985}. 
The $^{14}$N hyperfine splitting can be resolved easily at low values of $J$ 
\cite{VyCN_Sutter_1985,VyCN_Colmont_1997}. It can be resolved at least in part 
for higher-$J$ values in $a$-type $R$-branch transitions when $K_a$ 
approaches $J$ \cite{VyCN_Baskakov_1996}, see also Fig.~\ref{VyCN_lab-fig}. 
In this context, it is interesting to note that Gerry and 
Winnewisser \cite{VyCN_Gerry_1973}
published two candidate lines for the $J = 14 - 13$, $K_a = 13$ transition 
at 133180.473 and 133181.018~MHz. 
In fact, both lines belong to this transition; the former is the $F = J$ 
hyperfine component, whereas the latter is the overlap of the 
$F = J \pm 1$ hyperfine components. In those days, $\sim$35 years ago, 
it was probably neither expected nor feasible to calculate hyperfine splitting 
at such comparatively high quantum numbers.

\begin{figure}[ht]
   \begin{center}
   \includegraphics[width=12cm]{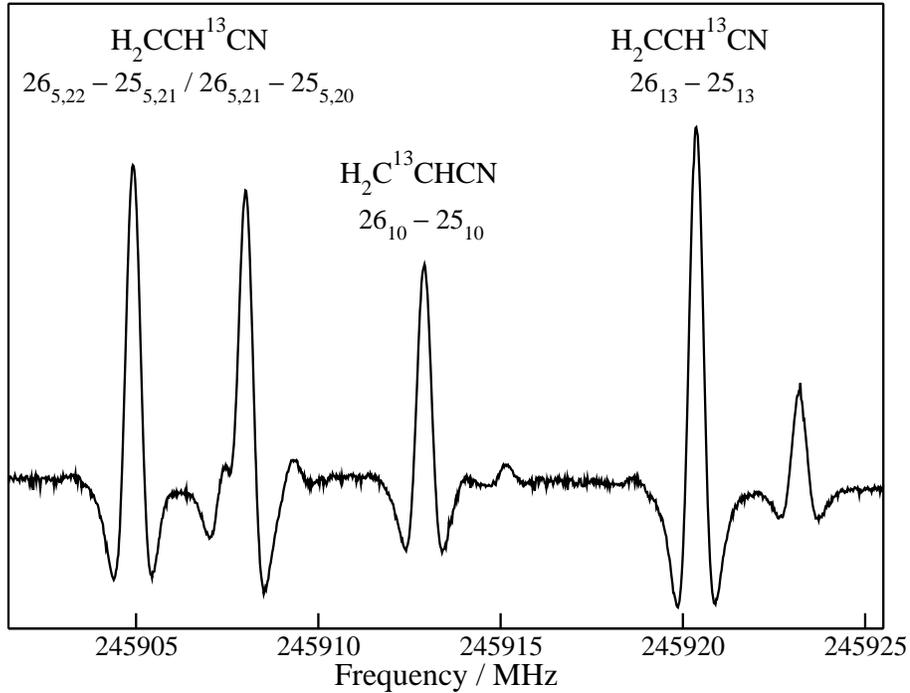}
   \caption{Portion of the rotational spectrum of vinyl cyanide in the region of the 
            $J = 26 - 25$ $a$-type $R$-branch transitions of H$_2$C=CH$-^{13}$C$\equiv$N 
            and H$_2$C=$^{13}$CH$-$C$\equiv$N. The stronger features in this recording 
            can be assigned to these two isotopic species. Since $K_a + K_c = J$ or $J + 1$, 
            $K_c$ has been omitted for the prolate paired transitions having $K_a = 10$ and 
            13, respectively. The weaker features can not be assigned at present. 
            As one can see, two weaker lines affect the position of the $26_{5,21} - 25_{5,20}$ 
            transition of H$_2$C=CH$-^{13}$C$\equiv$N.
            \label{VyCN_Iso_1mm}}
   \end{center}
\end{figure}

Selected $a$-type $R$-branch transitions of the singly substituted 
$^{13}$C and $^{15}$N isotopic species of VyCN were readily found 
in the 4 and 3~mm regions based on the previous data. $^{14}$N hyperfine 
splitting was resolved in part for transitions with high $K_a$ values 
as shown in Fig.~\ref{VyCN_lab-fig}. 
Subsequently, selected transitions were searched for at $239 - 253$~GHz,  
around 320~GHz, and near 350~GHz.
No $b$-type transitions were searched for specifically at lower frequencies 
as these were not only weak, but also their uncertainties were comparatively 
large initially. 
Nevertheless, some $b$-type transitions could be identified for all the 
$^{13}$C-containing species in wider scans around the $a$-type $R$-branch transitions 
of the main isotopic species recorded in the 4 and 3~mm regions. 
The measurements at higher frequencies improved the predictions for these 
assignments. Moreover, one $K_a = 1 - 0$ transition each was found for 
H$_2$C$^{13}$CHCN and for H$_2$CCH$^{13}$CN near 245 and 250~GHz, respectively. 
One should keep in mind that even though the $b$-type transitions of 
these isotopic species are considerably stronger at these higher frequencies, 
those of the main isotopic species in the molecule's ground and 
low-lying excited states are stronger, too. 
Moreover, the number of $a$-type $R$-branch transitions increases with $J$. 
Overall, this results in a strongly increased probability for overlap 
of two lines or for perturbations of the apparent line positions caused 
by the proximity of two lines as can be seen in Fig.~\ref{VyCN_Iso_1mm}.

The maximum $J$ values were 35 to 38 for the minor isotopic species, 
and the $K_a$ values reached 17 to 20.

Selected transitions were recorded for the main isotopic species 
in similar frequency regions. The transitions were easily found because 
of the extensive data set from previous laboratory studies
\cite{VyCN_Gerry_1973,VyCN_Demaison_1973,VyCN_Sutter_1985,VyCN_Cazzoli_1988,VyCN_Demaison_1994,VyCN_Baskakov_1996,VyCN_Colmont_1997}.
The maximum $J$ and $K_a$ values were 80 and 30, respectively.

The data obtained in the course of the present investigation were fit together 
with previous data. Uncertainties of 100~kHz were given for the microwave transitions 
from Refs.~\cite{VyCN_Gerry_1973,VyCN_Demaison_1973} as stated in the former. 
Average frequencies were used for the few lines that were reported in both studies, 
and the uncertainties were set to 70~kHz. Millimeter-wave transitions from 
\cite{VyCN_Gerry_1973} quoted to 10~kHz digits were given 150~kHz uncertainties as stated; 
those quoted to 1~kHz digits were reported to be more accurate by more than a factor of 10, 
suggesting uncertainties of 10~kHz. While this seemed to be reasonable for essentially 
all transitions not affected by partially resolved hyperfine structure, 
20~kHz appeared to be more appropriate for those that were affected. 
The reported uncertainties were largely used for the experimental lines from 
Refs.~\cite{VyCN_Sutter_1985,VyCN_Demaison_1994,VyCN_Colmont_1997}; 50 and 5~kHz, 
respectively, were ascribed to the data from 
Refs.~\cite{VyCN_Cazzoli_1988,VyCN_Baskakov_1996} on the basis of the 
average residuals between observed and calculated frequencies in these two papers.
Very few transitions were omitted from the final fits because of large residuals. 
These included the weak $F = 1 - 0$ hyperfine components of the 
$J_{K_a,K_c} = 2_{1,2} - 1_{1,1}$ transition for H$_2^{13}$CCHCN and H$_2$C$^{13}$CHCN 
\cite{VyCN_Colmont_1997} because trial fits with these lines weighted out not only 
increased the residuals of these two lines by a factor of about two 
but also improved the quality of the fit. 

The spectroscopic parameters determined for the main isotopic species are essentially 
the same as in Ref.~\cite{VA_2004}. Two additional decic and one dodecic parameters 
were necessary to fit the high-$K_a$ $a$-type $R$-branch transitions.
Initial parameters for the minor isotopic species were taken from 
Ref.~\cite{VyCN_Colmont_1997} and transformed to Watson's $S$-reduction. 
Higher order parameters were taken from 
the main isotopic species and scaled by the appropriate powers of the ratios of 
$2A - B - C$, $B + C$, and $B - C$. As observed previously \cite{VyCN_Colmont_1997}, 
$D_K$ did not follow this trend. Thus, $H_K$ and $L_K$ were derived 
similarly from the ratios of $D_K$. 
The resulting spectroscopic parameters are given in Table~\ref{spec_parameters}.

The experimental transition frequencies with assignments and uncertainties as well as 
residuals between observed frequency and those calculated from the final set of 
spectroscopic parameters are available as supplementary material (Table~S1). 
The detailed fit files as well as related files will be available 
in the Cologne Database for Molecular Spectroscopy, 
CDMS\footnote{http://www.astro.uni-koeln.de/vorhersagen/; short-cut: 
http://www.cdms.de/}~\cite{CDMS1,CDMS2}, as will be extensive predictions 
of the rotational spectra.

\section{Radioastronomical observations}
\label{obs-astro}

We carried out a complete line survey in the 3\,mm atmospheric window between 
80 and 116 GHz toward the hot core region Sagittarius B2(N)
(hereafter Sgr~B2(N) for short). Additional spectra were also obtained 
in the 2 and 1.3\,mm windows. The observations were performed in January 2004, 
September 2004, and January 2005  with the IRAM 30m telescope on Pico Veleta, 
Spain. Details about the observational setup and the data reduction are given 
in \cite{AAN-det}. 
An rms noise level of 15--20~mK on the $T^\star_{\mathrm{a}}$ scale 
was achieved below 100~GHz, 20--30~mK between 100 and 114.5~GHz, and 
about 50~mK between 114.5 and 116~GHz. 

The overall goal of our survey was to characterize the molecular content of Sgr~B2(N). 
It also allows searches for new species once all the lines emitted by 
known molecules have been identified, including vibrationally and 
torsionally excited states. We detected about 3700 lines above $3\sigma$ 
over the whole 3\,mm band.
These numbers correspond to an average line density of about 100 features 
per GHz. Given this high line density, the assignment of a line to a 
given  molecule can be trusted only if all lines emitted by this molecule in 
our frequency coverage are detected with the right intensity and no predicted
line is missing in the observed spectrum. The XCLASS software (see 
\cite{Orion-KL_795-903}) was used to model the emission of all known molecules 
in the local thermodynamical equilibrium approximation (LTE for short). 
More details about this analysis are given in \cite{AAN-det}. 
So far, we have identified 51 different molecules, 60 
isotopologues, and 41 vibrationally/torsionally excited states in Sgr~B2(N). 
This represents about $60\%$ of the lines detected above the $3\sigma$ level. 

The frequency ranges covered at 3 and 2\,mm contain several hundred 
transitions of the three $^{13}$C isotopologues of VyCN. The LTE 
modeling shows, however, that the transitions with the line strength times the 
appropriate dipole moment, $S\mu^2$, smaller than 40 D$^2$ are much too weak 
to be detectable with the sensitivity we achieved and the physical parameters
we explored. Out of 291 transitions above this $S\mu^2$ threshold, 194 are 
detected in our line survey toward Sgr~B2(N), either well isolated or in 
groups, or somewhat blended with emission from other molecules. The remaining 
97 transitions are heavily blended with emission from other molecules or 
weaker than the noise level, and cannot be identified in our spectrum. 
The detected transitions are listed in Table~S2 of the supplementary material. 
To save some space, when two transitions have a frequency difference 
smaller than 0.1 MHz which is not resolved in our astronomical data, 
only the first one is listed. The transitions are numbered in Col.~1, 
and their quantum numbers are given in Col.~3. The identities of the $^{13}$C 
isotopologues are coded in Col.~2. The frequencies, the frequency 
uncertainties, the energies of the lower levels in temperature units, and the 
$S\mu^2$ values are listed in Col.~4, 5, 6, and 7, respectively. 
Since the spectra are in most cases close to the line confusion limit 
and it is difficult to measure the noise level, we give in Col.~8 
the rms sensitivity computed from the system temperature 
and the integration time (see \cite{AAN-det} for details). 
The isolated transitions or groups of transitions that could be easily 
identified are labeled ``no blend'' in Col.~13, while for the other 
transitions we indicate which molecule contaminates the observed line. 
We identified the lines of the $^{13}$C isotopologues of VyCN and the
blends affecting them with the LTE model of these three molecules 
and the LTE model including all molecules identified in our survey thus far. 
Table~\ref{t:vycn13c30mmodel} shows the parameters of our best-fit LTE model 
of the $^{13}$C isotopologues of VyCN as well as the parameters of our 
best-fit LTE model of VyCN itself.

\begin{figure}[ht]
   \begin{center}
   \includegraphics[width=9cm,angle=270]{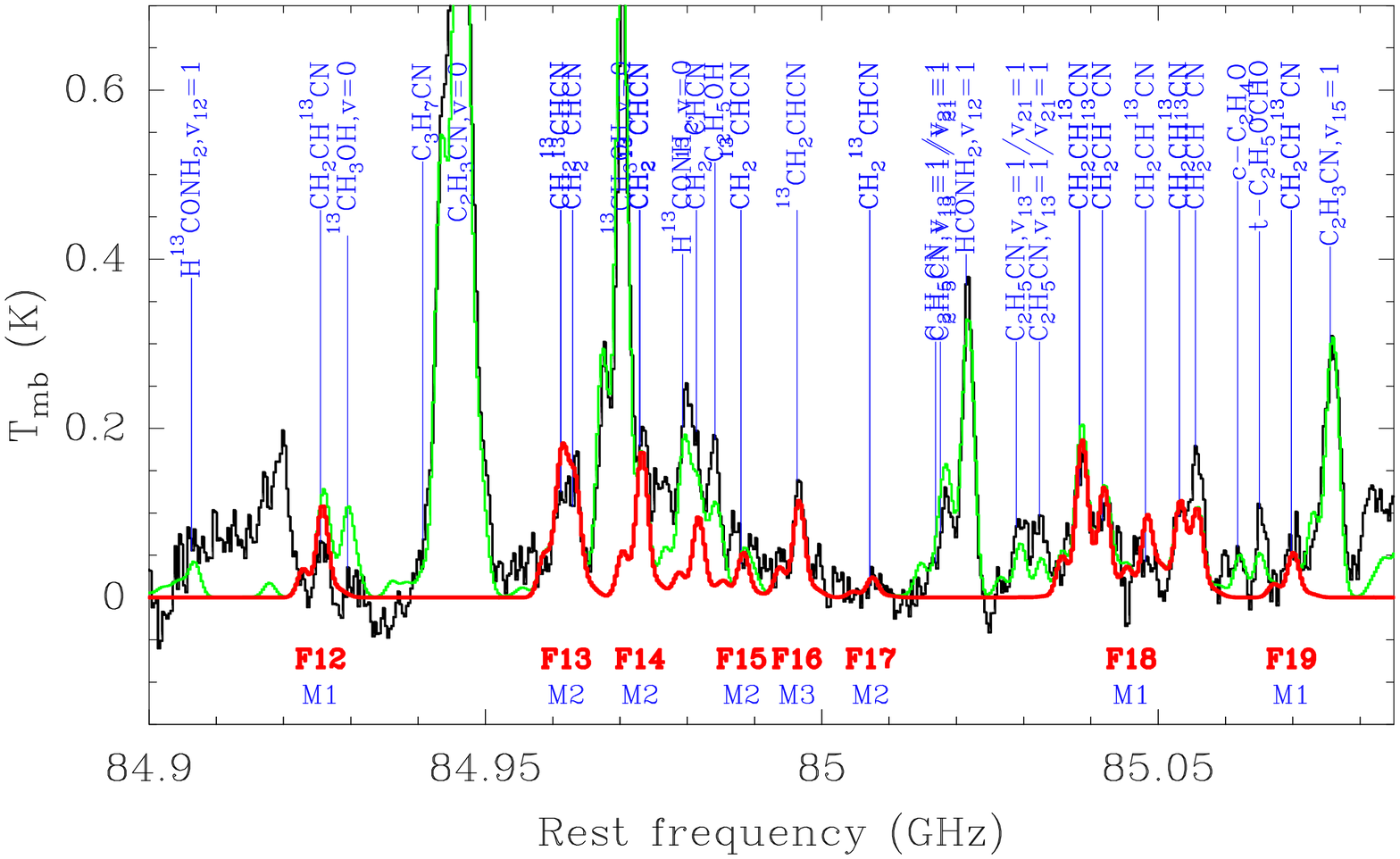}
   \includegraphics[width=9cm,angle=270]{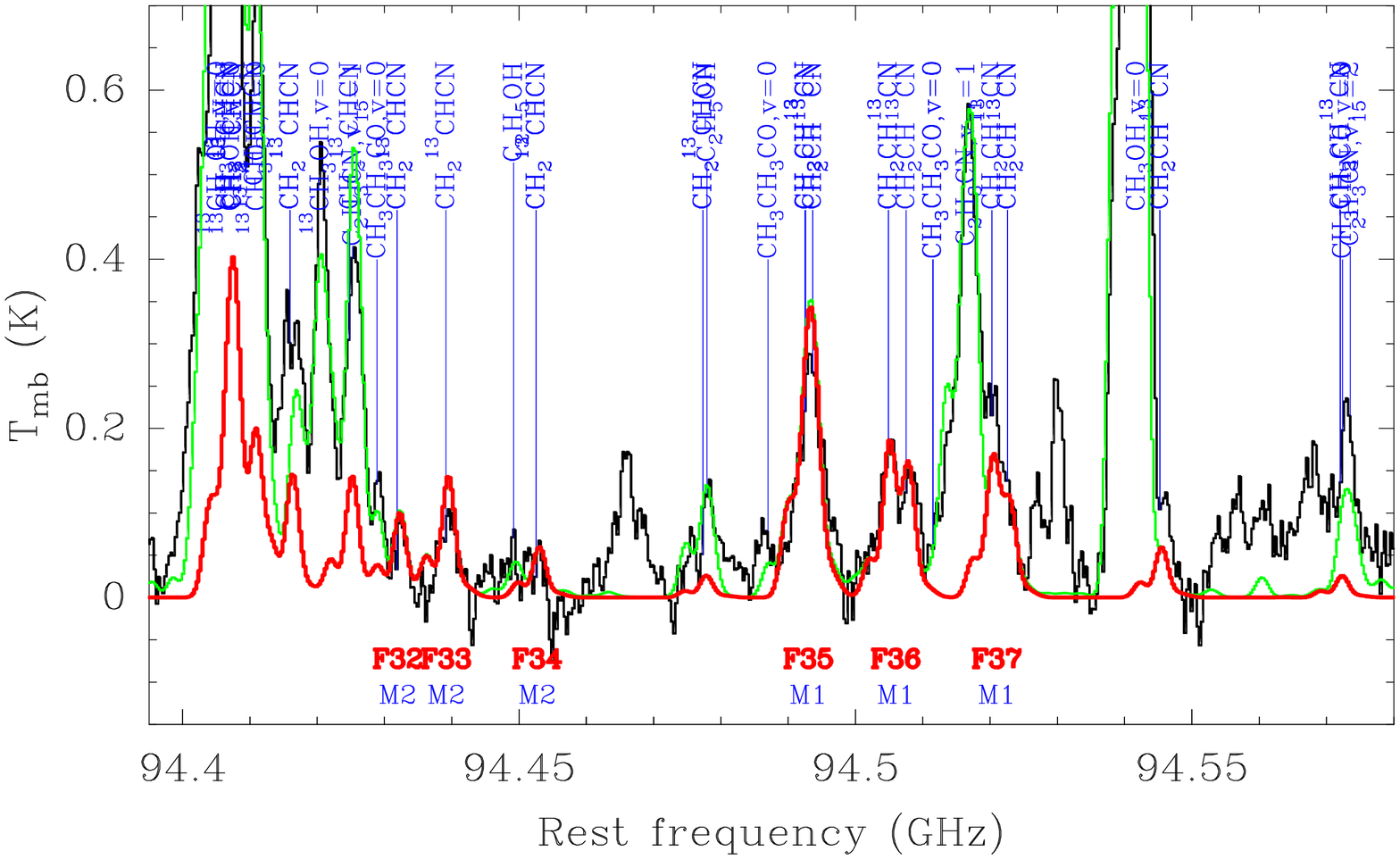}
   \caption{Selected transitions of the $^{13}$C isotopologues of vinyl 
            cyanide detected with the IRAM 30m telescope. In each panel, the 
            spectrum observed toward Sgr~B2(N) is shown in black, while the 
            LTE synthetic spectrum including the three $^{13}$C isotopologues 
            of vinyl cyanide is overlaid in red and the LTE model including all
            identified molecules in green. The detected features are labeled 
            in red, as listed in Col.~9 of Table~S2 in the supplementary material.
            The identified lines are labeled in blue (M1, M2, and M3 stand for 
            the three $^{13}$C isotopologues as in Col. 2 of Table~S2). 
            All observed lines which 
            have no counterpart in the green spectrum are still unidentified. 
            The systemic velocity assumed for Sgr~B2(N) is 64 km~s$^{-1}$.}
            \label{VyCN_obs-fig}
   \end{center}
\end{figure}

The detected transitions are grouped into ``features'' as listed in Col.~9 of
Table~S2 of the supplementary material, a feature being an observed line 
containing one or more (overlapping) transitions of the $^{13}$C isotopologues 
of VyCN. There are 89 features detected toward Sgr~B2(N), 10 of which 
look relatively free of contamination and an additional 16 are only slightly 
blended with emission from other molecules. The remaining 63 features are 
blended with emission from other molecules but with the $^{13}$C isotopologues 
of VyCN contributing significantly (at least $\sim 50\%$) to the peak 
intensity according to our LTE modeling. The integrated intensities of all 
detected features are given in Col.~10, along with the integrated intensities 
of our best-fit model of the $^{13}$C isotopologues of VyCN (Col.~11) 
and of the best-fit model including all molecules (Col.~12). Two selected frequency 
ranges containing Features F12 to F19 and F32 to F37 are shown in 
Fig.~\ref{VyCN_obs-fig}. The synthesized spectrum corresponding to the three 
$^{13}$C isotopologues of VyCN is plotted in red, while the model 
including all molecules is overlaid in green.

We derived the model parameters shown in Table~\ref{t:vycn13c30mmodel} from
all the constraints provided by the transitions of the $^{13}$C isotopologues 
but also from the main isotopic species which we easily detect in our survey 
in the ground state and in several vibrationally excited states (Belloche et al., 
\textit{in prep.}, see also \cite{VyCN-vib}). The strongest velocity component 
is detected at 63~km~s$^{-1}$ and corresponds to the so-called 
``Large Molecule Heimat'' (see \cite{AAN-det}). 
The second velocity component at 73~km~s$^{-1}$ is weaker, but still obvious 
in many transitions of the main isotopologue in the ground state and 
in the $v_{11} = 1$ and $v_{15} = 1$ vibrationally excited states. 
It is also detected in many transitions of the $^{13}$C isotopologues 
(e.g. in Feature 16 of Fig.~\ref{VyCN_obs-fig}). 
The third (wing) component at 55~km~s$^{-1}$ is much weaker. 
It is well seen in many transitions of the main isotopic species of 
vinyl cyanide and is (barely) detected in Feature 44 of $^{13}$CH$_2$CHCN, 
but it is too weak to be detected in all other $^{13}$C isotopologue transitions 
covered by our survey.
The maximum opacity of the $^{13}$C isotopologue transitions is 0.5 in our LTE model. 
Therefore these transitions are optically thin and the source size 
and column density are degenerated. On the other hand many transitions of 
the main isotopic species are optically thick, which allowed to derive the source size, 
once a good estimate of the temperature had been obtained. 
The same source size was used for all $^{13}$C isotopologues. 
Finally, no attempts were made to fit the transitions in the 1.3~mm window 
since the line confusion is much more severe and the position of the baselines 
much more uncertain. In addition, our LTE model predicts emission line
intensities stronger than observed for many molecules in this window. 
Apart from the uncertain baselines, the level of which is likely to 
be overestimated because of line confusion, we suspect that the dust emission 
is optically thick in this window and affects significantly the line emission 
from the compact hot cores, but we have not tried to model this effect so far. 

Finally, we searched for C$_2$H$_3$C$^{15}$N in our survey toward Sgr~B2(N) but
did not detect this isotopologue. 
Using the same source size, temperature, and linewidth as for the $^{13}$C isotopologues, 
a column-density upper limit of $3 \times 10^{15}$ cm$^{-2}$ could be derived 
in the LTE approximation. 
From a tentative detection of the $^{15}$N isotopologue of cyanoacetylene 
HC$_3$N in our survey toward Sgr~B2(N) (Belloche et al., \textit{in prep.}), 
we derive a $^{13}$C$^{14}$N/$^{12}$C$^{15}$N ratio of about 15, while Wilson 
and Rood \cite{Wilson94} mention a ratio larger than 30 for the Galactic 
Center region. In any case, the column-density upper limit derived here for 
C$_2$H$_3$C$^{15}$N is consistent with both values.

\section{Discussion of laboratory spectroscopic investigations}
\label{lab-disc}

The hyperfine parameters are largely determined by the data published 
in Ref.~\cite{VyCN_Colmont_1997}; additional information obtained in the 
present work for the isotopic species with $^{13}$C or previously 
for the main isotopologue \cite{VyCN_Gerry_1973,VyCN_Sutter_1985,VyCN_Baskakov_1996} 
have very small effects. Omission of one hyperfine component each for 
H$_2^{13}$CCHCN and H$_2$C$^{13}$CHCN actually increased their uncertainties of 
$\chi _{bb}$ and $\chi _{cc}$ slightly and changed their values by an amount 
slightly outside the combined uncertainties. 

The mechanical parameters for the main isotopic species agree  
very well with those determined in a recent combined fit \cite{VA_2004}. 
The uncertainties in the present work are somewhat smaller despite using 
three additional distortion parameters. These parameters are of the right order of 
magnitude when compared with the lower order parameters.
Three of the off-diagonal octic distortion terms ($l_2$, $l_3$, and $l_4$) 
are barely determined. 
Nevertheless, they were retained in the final fit as their values seemed 
to be of the correct order of magnitude. Moreover, omitting any one of 
these parameters deteriorated the fit slightly and had only modest effects 
on the remaining parameters. The values of the higher order distortion parameters 
should be viewed with some caution as additional experimental transition frequencies 
and the inclusion of still higher order terms may affect their values somewhat outside 
the reported uncertainties.

Considerably more distortion parameters have been determined for the 
minor isotopic species in the present work compared with the previous study 
of Colmont et al. \cite{VyCN_Colmont_1997}. 
Because of the change in reduction a direct comparison is not always possible. 
However, $\Delta _K$ and $D_K$ are very similar in a near-prolate top as 
VyCN. Their estimates of the former agree very well with the presently 
determined $D_K$ value, even in the case of the $^{15}$N species for which 
we did not observe $b$-type transitions. The distortion parameters of the 
minor isotopic species are very close to those of the main isotopic species 
as one would expect because the rotational constants are also quite similar.
The use of higher order parameters transfered from the main isotopic species 
and kept fixed in the analysis may seem excessive at first sight, in particular 
if one keeps in mind that some of these are comparatively uncertain. 
However, most of the fits for the minor isotopic species were performed without 
the parameters higher than those of octic order (the $L$s). In those fits, 
the values for $L_{KKJ}$ were much smaller than that of the main isotopic species 
(by about a factor of 1.5) whereas those for $L_{JK}$ were much larger in magnitude 
than that of the main isotopic species (by about a factor of $1.5 - 2$).
With estimates for these high-order parameters included in the fit the values 
for $L_{KKJ}$ and $L_{JK}$ changed only barely from those of the main isotopologue 
if the uncertainties are taken into account.

\section{Discussion of the radioastronomical detection}
\label{astro-disc}

We have detected the three $^{13}$C isotopologues of vinyl cyanide toward 
Sgr~B2(N). Although many transitions are blended with emission from other 
molecules, 26 features have been found completely free of contamination or
only slightly contaminated. The emission arises from several velocity 
components. The two velocity components at 63 and 73 km~s$^{-1}$ were also 
detected by Belloche et al. \cite{AAN-det} with the IRAM Plateau de Bure 
interferometer (PdBI) and the Australia Telescope Compact Array (ATCA) in 
ethyl cyanide C$_2$H$_5$CN, cyanoacetylene HC$_3$N and its isotopologue 
HC$^{13}$CCN in their $v_7 = 1$ excited states, and methanol CH$_3$OH in its 
excited state $v_t = 1$. The small source size derived here (2.3$''$) is
consistent with the size of the VyCN emission detected with BIMA by 
Liu and Snyder \cite{VyCN_Liu_1999} (see their figure 2a) and with the 
similar fluxes detected by Friedel et al. \cite{Friedel04} with BIMA and 
the NRAO 12\,m telescope (only 10$\%$ of the 12\,m flux is missed by BIMA). 
This source size is also similar to the one measured for aminoacetonitrile 
\cite{AAN-det}. Therefore, the emission of VyCN and its $^{13}$C 
isotopologues arises mainly from the two compact hot cores separated by about 
5.3$''$ in the North-South direction. These two hot cores are not spatially 
resolved with the IRAM 30m telescope but are clearly separated with 
interferometers like the PdBI and the ATCA. 
The main one with a velocity of 63~km~s$^{-1}$ corresponds to the
so-called ``Large Molecule Heimat'' (LMH for short).
The third velocity component, identified at 55 km~s$^{-1}$, most likely 
corresponds to the blueshifted linewing detected with the PdBI in 
cyanoacetylene HC$_3$N (see Fig. 5m of \cite{AAN-det}), and may be 
associated with an outflow. 

We derived a column density of $8 \times 10^{17}$ cm$^{-2}$ for the main 
velocity component of VyCN which corresponds to the LMH (see 
Table~\ref{t:vycn13c30mmodel}).
This value is a factor of 3-4 larger than the one calculated by Liu and 
Snyder \cite{VyCN_Liu_1999}, which is not surprising since they analyzed 
their spectrum of VyCN at 83.2~GHz assuming optically thin emission 
whereas our LTE modeling indicates an optical depth of $\sim$2.3 for 
this transition in our survey. In addition, they may not have taken into 
account the contribution of the low-lying vibrational modes to the partition
function as was done here (about 25$\%$). The 
temperature derived from our LTE analysis (170~K) is in good agreement 
with the temperature measured by Ikeda et al. \cite{Ikeda01} for their 
VyCN transitions observed with the Nobeyama radiotelescope ($141 \pm 56$~K). 
On the other hand, no evidence for a high-temperature component (440~K) 
was found for the main isotopologue in its ground vibrational state 
in the present survey in contrast to what was reported by Nummelin and 
Bergman \cite{VyCN-vib}. Since the column densities of VyCN and its $^{13}$C
isotopologues have been independently measured here, the $^{12}$C/$^{13}$C
isotopic ratio of VyCN can be calculated from the column-density ratio. 
As a result, we find a $^{12}$C/$^{13}$C isotopic ratio of 21 for VyCN, 
which is fully consistent with the standard value listed by 
Wilson and Rood for the Galactic Center \cite{Wilson94}. 

With a mean H$_2$ column density of 
$N_{\mathrm{H}_2} = 1.3 \times 10^{25}$~cm$^{-2}$ \cite{AAN-det}, 
a relative abundance of $\sim 6.2 \times 10^{-8}$ is derived for VyCN in the LMH, 
and $\sim 2.9 \times 10^{-9}$ for each of its $^{13}$C isotopologues. 
The abundance of VyCN derived here is more than one order of 
magnitude larger than the abundances recently found by Fontani et al. in a 
sample of massive hot cores \cite{Fontani07}. 

We also detect the $^{13}$C isotopologues of ethyl cyanide (C$_2$H$_5$CN, EtCN 
for short) in our survey of Sgr~B2(N). Rest frequencies were taken from the 
CDMS~\cite{CDMS1,CDMS2}; the entries were based on the laboratory data  
from Ref.~\cite{det_13C_EtCN}. The latter work also reports the detection 
of $^{13}$C isotopologues of EtCN for the first time in space.
The current LTE analysis of our Sgr~B2(N) survey (Belloche et al., 
\textit{in prep.}) shows that the detected lines are optically thin, with 
opacities smaller than 0.4. 
For the main velocity component of each $^{13}$C 
isotopologue, a column density of $4 \times 10^{16}$~cm$^{-2}$ is found with 
a source size of 3$''$ derived from the detected EtCN lines and a 
temperature of 170~K. In the LTE approximation, the contribution of the 
low-lying vibrational modes to the partition function is estimated to be about 
40$\%$ at 170 K, which yields a total column density of 
$5.6 \times 10^{16}$~cm$^{-2}$ for the $^{13}$C isotopologues of EtCN. 
Therefore, the column-density ratio of the $^{13}$C isotopologues of VyCN to 
the $^{13}$C isotopologues of EtCN is $ R \approx 0.7$ in the LMH, which is in 
good agreement with what Liu and Snyder \cite{VyCN_Liu_1999} found for the 
$^{12}$C species. However, they may not trace the same material in this source 
since the VyCN emission looks single-peaked in interferometric maps while the 
EtCN emission is definitively double-peaked (see \cite{VyCN_Liu_1999} and 
Fig.~5h of \cite{AAN-det}). For the second velocity component corresponding to 
the northern hot core in Sgr~B2(N), we find a column density of 
$5 \times 10^{16}$~cm$^{-2}$ for the $^{13}$C isotopologues of EtCN with a 
source size of 2.3$''$ and a temperature of 170~K. After correction for the
contribution of the low-lying vibrational modes, the total column density is
about $7 \times 10^{16}$~cm$^{-2}$. This gives a column-density ratio 
$R \approx 0.16$ for the northern hot core of Sgr~B2(N). 

In hot cores, EtCN is thought to evaporate from the dust grains where it was 
formed and to subsequently form VyCN through gas-phase reactions 
\cite{Fontani07,Caselli93}. In the dynamical-chemical calculations of Caselli 
et al. \cite{Fontani07,Caselli93} modeling the gas phase and surface chemistry 
in a collapsing core with a density of 10$^7$ cm$^{-3}$ and a temperature of 
200~K, the abundance ratio $R$ of VyCN to EtCN increases initially with time 
as VyCN is progressively produced from EtCN, reaches a maximum of about 0.5, 
and then starts to decrease when the abundances of many complex 
molecules also drop. This ratio $R$ may therefore be used as a chemical clock 
\cite{Fontani07}. The ratio $R \approx 0.7$ found here for the LMH is even larger 
than the maximum predicted by the model. 
Fontani et al. \cite{Fontani07} mentioned that the model may underestimate 
the production of VyCN (thus $R$ too) due to a missing reaction of EtCN 
with H$_3$O$^+$ in the chemical network. If we take the large
value found in LMH as the true maximum, then the age of the LMH is 
$\sim 8 \times 10^4$~yr since the beginning of the evaporation of the grain 
mantles in the framework of this model. On the other hand, the ratio $R$ about
4 times smaller found for the northern hot core suggests that it is at a 
different evolutionary stage than the LMH, either younger ($3 \times 10^4$~yr) 
or older ($1.6 \times 10^5$~yr), depending on whether it is before or after 
the peak of $R$. We note however that the H$_2$ density in the LMH is about
$2 \times 10^8$ cm$^{-3}$ \cite{AAN-det}, which is one order of magnitude 
larger than the density assumed by Caselli et al. \cite{Caselli93} for their model. 
The chemical timescales may therefore be shorter in the LMH than in the model, 
and the ages determined above may be overestimated.

\section{Conclusion}
\label{concl}

Very accurate transition frequencies have been obtained for five 
isotopic species of VyCN in their ground vibrational states within 
the millimeter-wave domain. 
The resulting spectroscopic parameters are accurate enough 
to identify the minor isotopic species in the interstellar medium 
or circumstellar envelopes up to about 700~GHz. 
This corresponds approximately with the foreseen upper frequency limit 
of ALMA of 720~GHz which may even be extended up to around 950~GHz later. 
On the other hand, none of the high frequency molecular line surveys 
of Orion~KL carried out with comparatively large ($\sim$10~m) 
single dish telescopes covering the frequency regions $455 - 507$~GHz
\cite{Orion-KL_455-507}, $607 - 725$~GHz \cite{Orion-KL_607-725}, and 
$795 - 903$~GHz \cite{Orion-KL_795-903} revealed any features of the main 
isotopic species of VyCN while even vibrationally excited VyCN 
was observed in the $325 - 360$~GHz survey \cite{Orion-KL_325-360}. 
Therefore, it is not clear at present if the minor isotopic species 
of VyCN will be observable with ALMA in its highest frequency bands; 
however the expected very high sensitivity and spatial resolution 
may make this fairly likely. 
Because of previous laboratory data~\cite{VyCN_Demaison_1994}, 
predictions for the main isotopic species should be reliable up to 
at least 1~THz. The current data significantly improve the 
predictions for the weaker lines.
Extensive predictions of the rotational spectra will be available 
in the CDMS~\cite{CDMS1,CDMS2} for all these species.
These predictions will be a contribution to minimize the so-called 
``weed problem'', that means that features of abundant molecules 
in space, such as VyCN, may prevent weaker features of desired molecules 
to be identified unambiguously unless sufficiently weak features of the 
more abundant species are known accurately enough. 
The ``weed problem'' is of particular concern for radiotelescope arrays 
such as the SMA or, even more so, for ALMA.

With the accurate transition frequencies obtained in the present work, we 
could identify for the first time in space the three $^{13}$C species of 
VyCN in our 3\,mm molecular line survey of the massive star forming region 
Sgr~B2(N). The emission is detected in two compact ($\sim 2.3''$) hot cores
with a temperature of 170~K, a column density of $\sim 3.8 \times 10^{16}$ 
and $1.1 \times 10^{16}$ cm$^{-2}$, and a velocity of 63 and 73~km~s$^{-1}$,
respectively. In the strongest source, the ``Large Molecule Heimat'' 
with a velocity of 63~km~s$^{-1}$, a relative abundance of $2.9 \times 10^{-9}$ 
has been derived for each $^{13}$C species of VyCN, and an isotopic ratio 
$^{12}$C/$^{13}$C of 21 has been measured. Based on a comparison to the column 
densities measured for the $^{13}$C species of ethyl cyanide also detected in 
this survey, it is suggested that the two hot cores of Sgr~B2(N) are in 
different evolutionary stages. 

\section{Acknowledgments}
\label{Acknowledgments}

H.S.P.M. is grateful for initial funding provided by the Deutsche Forschungsgemeinschaft 
(DFG) through the Sonderforschungsbereich (SFB) 494. Recently, he has been supported by the 
Bundesministerium f\"ur Bildung und Forschung (BMBF) administered through Deutsches Zentrum 
f\"ur Luft- und Raumfahrt (DLR). His support was aimed in particular at maintaining the CDMS.
Further financial resources have been furnished by the Land Nordrhein-Westfalen.



\newpage
\begin{table*}
\caption{Spectroscopic parameters$^a$ (MHz) of isotopic species of 
vinyl cyanide.}
\label{spec_parameters}
\smallskip
{\scriptsize
\renewcommand{\arraystretch}{1.15}
\begin{tabular}{lr@{}lr@{}lr@{}lr@{}lr@{}l}
\hline
Parameter & \multicolumn{2}{c}{H$_2$CCHCN} & \multicolumn{2}{c}{H$_2^{13}$CCHCN} &
 \multicolumn{2}{c}{H$_2$C$^{13}$CHCN} &  \multicolumn{2}{c}{H$_2$CCH$^{13}$CN} & \multicolumn{2}{c}{H$_2$CCHC$^{15}$N} \\
\hline
$A$                        & 49850&.69674\,(20)  & 49195&.292\,(89)   & 48639&.693\,(31)   & 49799&.643\,(26)    & 49655&.880\,(173)  \\
$B$                        &  4971&.163651\,(24) &  4837&.58023\,(12) &  4948&.98019\,(13) &  4948&.36515\,(12)  &  4819&.66825\,(18) \\
$C$                        &  4513&.877260\,(25) &  4398&.24203\,(12) &  4485&.38246\,(13) &  4494&.65445\,(13)  &  4387&.04599\,(17) \\
$D_K \times 10^3$          &  2714&.880\,(19)    &  2710&.8\,(163)    &  2597&.1\,(59)     &  2630&.3\,(64)      &  2762&.\,(35)      \\
$D_{JK} \times 10^3$       & $-$85&.01578\,(79)  & $-$86&.3393\,(51)  & $-$79&.9149\,(30)  & $-$85&.4947\,(32)   & $-$82&.6057\,(48)  \\
$D_J \times 10^3$          &     2&.182401\,(32) &     2&.11922\,(12) &     2&.11940\,(9)  &     2&.16156\,(10)  &     2&.04954\,(20) \\
$d_1 \times 10^6$          &$-$456&.5241\,(90)   &$-$438&.491\,(112)  &$-$453&.328\,(123)  &$-$451&.389\,(150)   &$-$420&.381\,(174)  \\
$d_2 \times 10^6$          & $-$30&.8801\,(42)   & $-$28&.023\,(42)   & $-$31&.835\,(32)   & $-$29&.678\,(41)    & $-$28&.361\,(78)   \\
$H_K \times 10^6$          &   384&.14\,(45)     &   384&.            &   364&.            &   370&.             &   384&.            \\
$H_{KJ} \times 10^6$       &  $-$6&.9401\,(81)   &  $-$7&.037\,(35)   &  $-$6&.785\,(18)   &  $-$6&.766\,(23)    &  $-$6&.931\,(51)   \\
$H_{JK} \times 10^6$       &  $-$0&.28487\,(27)  &  $-$0&.2835\,(37)  &  $-$0&.2607\,(16)  &  $-$0&.2912\,(19)   &  $-$0&.2633\,(34)  \\
$H_J \times 10^9$          &     5&.7252\,(92)   &     5&.539\,(42)   &     5&.381\,(49)   &     5&.612\,(66)    &     5&.073\,(67)   \\
$h_1 \times 10^{12}$       &  2276&.3\,(53)      &  2243&.\,(46)      &  2436&.\,(64)      &  2181&.\,(88)       &  1946&.\,(77)      \\
$h_2 \times 10^{12}$       &   367&.8\,(38)      &   330&.2           &   375&.6           &   360&.0            &   319&.2           \\
$h_3 \times 10^{12}$       &    89&.15\,(83)     &    79&.1           &    92&.9           &    87&.1            &    75&.5           \\
$L_K \times 10^9$          & $-$57&.4\,(33)      & $-$58&.            & $-$54&.7           & $-$55&.5            & $-$58&.0           \\
$L_{KKJ} \times 10^9$      &     1&.213\,(24)    &     1&.195\,(17)   &     1&.221\,(27)   &     1&.197\,(45)    &     1&.367\,(108)  \\
$L_{JK} \times 10^{12}$    & $-$81&.6\,(25)      & $-$57&.1\,(252)    & $-$76&.9\,(65)     & $-$82&.5\,(79)      & $-$81&.9\,(269)    \\
$L_{JJK} \times 10^{12}$   &     1&.785\,(37)    &     1&.63          &     1&.71          &     1&.76           &     1&.63          \\
$L_J \times 10^{15}$       & $-$21&.65\,(85)     & $-$19&.1           & $-$20&.8           & $-$21&.3            & $-$18&.9           \\
$l_1 \times 10^{15}$       &  $-$8&.89\,(103)    &  $-$7&.85          &  $-$8&.94          &  $-$8&.77           &  $-$7&.75          \\
$l_2 \times 10^{15}$       &  $-$3&.78\,(99)     &  $-$3&.23          &  $-$3&.75          &  $-$3&.60           &  $-$3&.11          \\
$l_3 \times 10^{15}$       &  $-$1&.42\,(38)     &  $-$1&.25          &  $-$1&.50          &  $-$1&.41           &  $-$1&.19          \\
$l_4 \times 10^{15}$       &  $-$0&.223\,(50)    &  $-$0&.186         &  $-$0&.230         &  $-$0&.211          &  $-$0&.174         \\
$P_{KKKJ} \times 10^{15}$  &$-$227&.\,(32)       &$-$210&.            &$-$202&.            &$-$225&.             &$-$218&.            \\
$P_{KKJ} \times 10^{15}$   &$-$108&.9\,(52)      & $-$99&.            & $-$99&.            &$-$107&.             &$-$102&.            \\
$S_{KKKKJ} \times 10^{18}$ &   157&.\,(17)       &   144&.            &   136&.            &   155&.             &   150&.            \\
$\chi_{aa}$                &  $-$3&.78907\,(40)  &  $-$3&.76634\,(55) &  $-$3&.80158\,(55) &  $-$3&.78580\,(40)  &      &n.\,a.       \\
$\chi_{bb}$                &     1&.68606\,(43)  &     1&.6547\,(41)  &     1&.6902\,(41)  &     1&.68435\,(167) &      &n.\,a.       \\
$\chi_{cc}$                &     2&.10301\,(49)  &     2&.1116\,(40)  &     2&.1114\,(40)  &     2&.10145\,(166) &      &n.\,a.       \\
$C_{aa} \times 10^3$       &     2&.20\,(30)     &     2&.20          &     2&.20          &     2&.20           &      &$-$          \\
$C_{bb} \times 10^3$       &     0&.78\,(17)     &     0&.78          &     0&.78          &     0&.78           &      &$-$          \\
$C_{cc} \times 10^3$       &     1&.43\,(17)     &     1&.43          &     1&.43          &     1&.43           &      &$-$          \\
\hline
\end{tabular}\\[2pt]
}
$^a$ Numbers in parentheses are one standard deviation in units of the least significant figures. 
Parameter values with no uncertainties given were estimated from those of the main isotopic species and kept fixed; 
see section~\ref{obs}. A long dash indicates parameters that are determinable in theory, but have not been determined 
here; n.\,a. stands for not applicable.\\
\end{table*}




\newpage
\begin{table}
 \centering
 \caption{
 Parameters of our best-fit LTE model of vinyl cyanide (VyCN) and its $^{13}$C isotopologues. 
 We used the same parameters for all three $^{13}$C isotopologues. For each species, 
 each line corresponds to a different velocity component in Sgr~B2(N).
}
 \label{t:vycn13c30mmodel}
 \vspace*{0.0ex}
 \begin{tabular}{cccccc}
 \hline\hline
 \multicolumn{1}{c}{Molecule} & \multicolumn{1}{c}{Size$^{a}$} & \multicolumn{1}{c}{T$_{\mathrm{rot}}$} & \multicolumn{1}{c}{N$^{b}$} & \multicolumn{1}{c}{FWHM} & \multicolumn{1}{c}{V$_{\mathrm{off}}$$^{c}$} \\ 
  & \multicolumn{1}{c}{\scriptsize ($''$)} & \multicolumn{1}{c}{\scriptsize (K)} & \multicolumn{1}{c}{\scriptsize (cm$^{-2}$)} & \multicolumn{1}{c}{\scriptsize (km~s$^{-1}$)} & \multicolumn{1}{c}{\scriptsize (km~s$^{-1}$)} \\ 
 \multicolumn{1}{c}{(1)} & \multicolumn{1}{c}{(2)} & \multicolumn{1}{c}{(3)} & \multicolumn{1}{c}{(4)} & \multicolumn{1}{c}{(5)} & \multicolumn{1}{c}{(6)} \\ 
 \hline
VyCN & 2.3 &  170 & $ 8.00 \times 10^{17}$ & 7.0 & -1.0 \\  & 2.3 &  170 & $ 2.40 \times 10^{17}$ & 7.0 & 9.0 \\  & 2.3 &  170 & $ 1.00 \times 10^{17}$ & 10.0 & -9.0 \\ \hline$^{13}$VyCN &  2.3 &  170 & $ 3.75 \times 10^{16}$ & 7.0 & -1.0 \\  &  2.3 &  170 & $ 1.13 \times 10^{16}$ & 7.0 & 9.0 \\  &  2.3 &  170 & $ 0.47 \times 10^{16}$ & 10.0 & -9.0 \\  \hline
 \end{tabular}
 \begin{list}{}{}
 \item[$(a)$]{Source diameter (FWHM).}
 \item[$(b)$]{The column densities listed here for all isotopic species of VyCN were computed 
 with partition functions that take into account contributions from the low-lying vibrational modes 
 which amount to about 25$\%$ in the LTE approximation at 170~K.}
 \item[$(c)$]{Velocity offset with respect to the systemic velocity of Sgr~B2(N) 
  V$_{\mathrm{lsr}} = 64$ km~s$^{-1}$.}
 \end{list}
 \end{table}


\clearpage

\clearpage

\clearpage

\end{document}